\documentclass[aps, twocolumn, pra, 10pt]{revtex4-2}
\usepackage{graphicx}  
\usepackage{xcolor}

{}
\usepackage[T1]{fontenc}
\usepackage{bm}        
\usepackage{amsfonts}  
\usepackage{amsmath}   
\usepackage{amssymb}   
\usepackage{braket}
\graphicspath{{Images/}}
\usepackage[colorlinks=true, allcolors=blue]{hyperref}
\usepackage[normalem]{ulem}
\usepackage{comment}

\begin{document}
\title{3-phases Confusion Learning}
\author{Filippo Caleca}
\affiliation{%
ENS de Lyon, Université Lyon 1, CNRS, Laboratoire de Physique, F-69342 Lyon,France
}%
\author{Simone Tibaldi}
\address{Dipartimento di Fisica e Astronomia dell'Universit\`a di Bologna, I-40127 Bologna, Italy}
\affiliation{INFN, Sezione di Bologna, I-40127 Bologna, Italy}
\author{Elisa Ercolessi}
\address{Dipartimento di Fisica e Astronomia dell'Universit\`a di Bologna, I-40127 Bologna, Italy}
\affiliation{INFN, Sezione di Bologna, I-40127 Bologna, Italy}

\begin{abstract}  
The use of Neural Networks in quantum many-body theory has seen a formidable rise in recent years. Among the many possible applications, one surely is to make use of their pattern recognition power when dealing with the study of equilibrium phase diagram. Within this context, Learning by Confusion has emerged as an interesting, unbiased scheme. The idea behind it briefly consists in iteratively label numerical results in a random way and then train and test a Neural Network; while for a generic random labeling the Network displays low accuracy, the latter shall display a peak when data are divided into a correct, yet unknown way. Here, we propose a generalization of this confusion scheme for systems with more than two phases, for which it was originally proposed. Our construction simply relies on the use of a slightly different Neural Network: from a binary classificator we move to a ternary one, more suitable to detect systems exhibiting three phases. After introducing this construction, we test is onto the free and the interacting Kitaev chain and on the one-dimensional Extended Hubbard model, always finding results compatible with previous works. Our work opens the way to wider use of Learning by Confusion, showing once more the usefullness of Machine Learning to address quantum many-body problems. 
\end{abstract}

\maketitle 
\section{Introduction}

The exponential growth in computational resources needed to solve the quantum many-body problem often restricts access to exact solutions, with some notable exceptions~\cite{doi:10.1142/5552}. Consequently, numerical techniques are frequently employed to validate physical intuitions or make predictions. Among these techniques are Quantum Monte Carlo~\cite{Sandvik_2010}, Density Matrix Renormalization Group (DMRG)~\cite{Schollw_ck_2011}, and Variational Monte Carlo~\cite{Becca_Sorella_2017}. Recently, especially in phase diagram reconstruction, Machine Learning algorithms have proven valuable in confirming theoretical data~\cite{Tibaldi, ourpaper} and even suggesting unexplored phases~\cite{RodriguezNieva2019, Long2020, Scheurer2020, Che2020, Lustig2020, Lidiak2020, Long2023, Rem2019}.

Within this context, Learning by Confusion~\cite{van_Nieuwenburg_2017} has emerged as an interesting unsupervised technique which uses Neural Networks to find phase transition points in quantum many-body systems. The main idea behind it is to train a Neural Network multiple times with the same dataset labeled differently each time until a peak in accuracy is found. A random labeling is likely to produce lower accuracy because the network is trying to learn mistakenly assigned labels, while a correct, yet unknown, labeling is likely to produce a larger accuracy. This is the signal of having found an optimal way of separating the data and, in our context, found a phase transition.
Learning by Confusion was firstly tested to catch the topological transition in the free Kitaev chain, the thermal phase transition in the classical Ising model and the many-body localization transition in the random field Heisenberg chain~\cite{van_Nieuwenburg_2017}. Later, it was also successfully employed to detect first order phase transitions~\cite{Richter_Laskowska_2023}, transitions in frustrated magnetic models~\cite{CORTE2021110702}, nuclear liquid-gas transitions~\cite{PhysRevResearch.2.043202} and entanglement breakdown~\cite{Gavreev_2022}.

Although this method proved efficient at detecting phase transitions in an unsupervised way we find ourselves often constrained to look for multiple phase transitions. For this reason, in this paper we generalize
this formalism to systems with more possible phases. While confusion learning was first introduced to detect single phase transitions -- i.e.
with binary labellings -- and the original scheme has also been employed to study models with multiple phase transitions~\cite{PhysRevE.99.043308}, in this work we develop an intuitive generalization of this method that considers sections of phase diagrams that might present two phase transitions. We therefore refer to the original method as 2-phases Learning and to our extension as 3-phases Learning.

The paper is structured as follow. In section~\ref{sec:Confusion} we explain in detail how 2-phases and 3-phases learning work. In section~\ref{sec:Results} we apply 2-phases learning and our new method, 3-phases learning, to four non-trivial models in quantum many body physics: the Kitaev chain, in its free and interacting version, and the one-dimensional Extended Hubbard model with two different shoulder potentials.
Finally, in section~\ref{sec:Conclusions} we draw the main conclusions and possible outlooks of this work.

\begin{figure*}
    \centering
    \includegraphics[width = \linewidth]{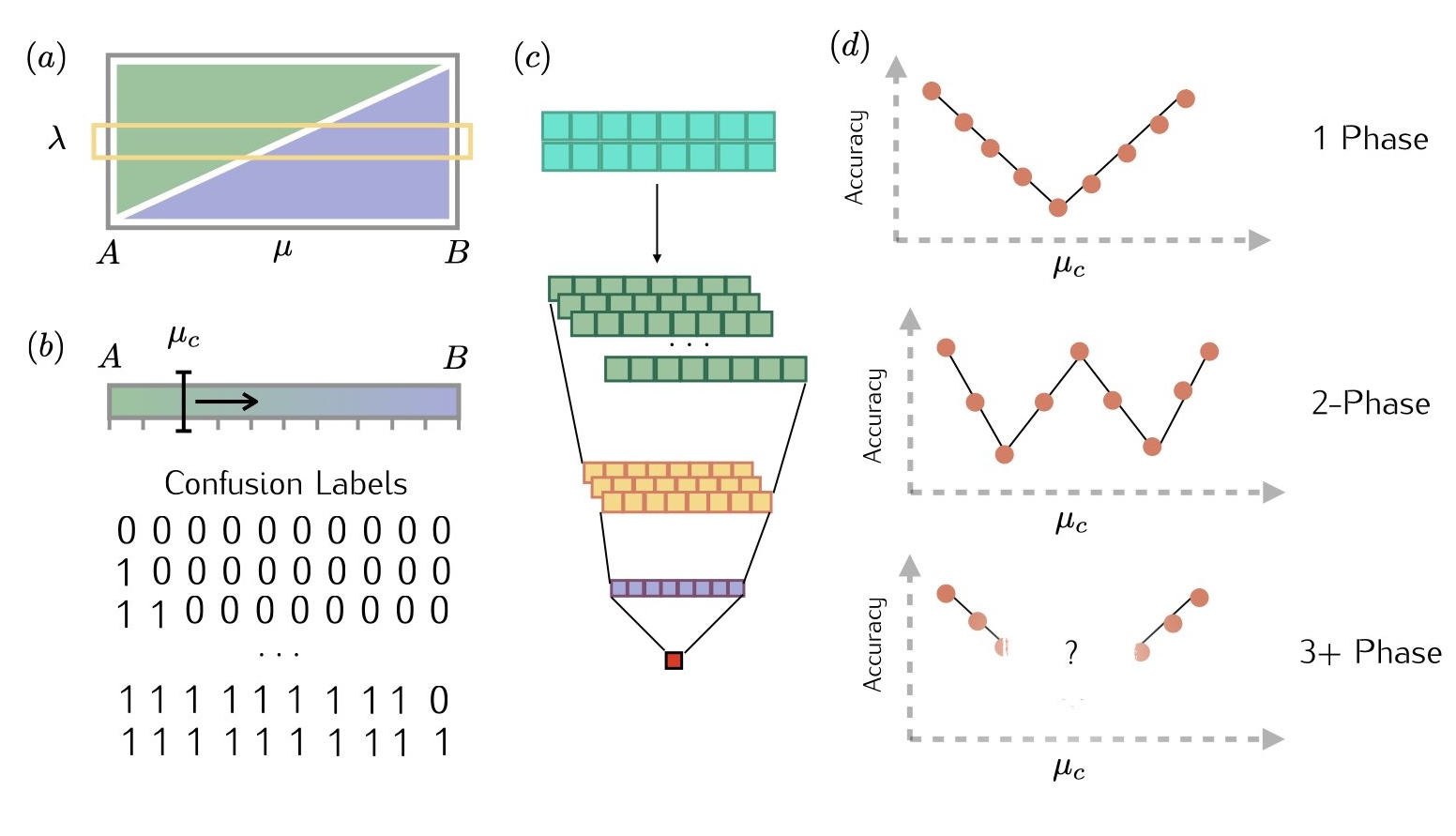}
    \caption{\textbf{Confusion Learning}. (a) We start by selecting a line of the phase diagram that may or may not cross a phase transition by fixing one parameter and changing the other one (in this example phase diagram $\lambda$ is fixed and $\mu$ is changed). (b) By sweeping a parameter $\mu_c$ in the discretized interval $[A, B]$ we generate different labellings for our data going from all zeros to all ones. (c) Scheme of the convolutional neural network used in the process. Blue is the input data; in green, yellow and purple the intermediate layers and finally the accuracy is the red square representing the output neuron. For each labelling listed below Confusion Labels we train a convolutional neural network and plot its accuracy. (d) We expect the canonical $V$-shape and $W$-shape in the case of no (top) or one (middle) phase transition. There is no case in literature for multiple phase transitions (bottom).}
    \label{fig:cflearning}
\end{figure*}

\section{Confusion Learning}
\label{sec:Confusion}

Learning by Confusion is a widely-known and successful unsupervised method to determine phase transition points \cite{https://doi.org/10.48550/arxiv.2204.04198,
	richterlaskowska2022learning, van_Nieuwenburg_2017}.
The main idea is to use a neural network to determine the phase transition point of a quantum system by scrambling the dataset until a good performance is reached.
The data used in our case is made of  observables (which should signal the presence of a phase transition) computed for different phase diagram points. Since we arrange them in a matrix form the network we consider in this work is a Convolutional Neural Networks (CNN) which are largely used in pattern recognition problems 
~\cite{SunProceedings}. A full description of the structure and functioning of CNNs is beyond our scope and we address the interested reader to~\cite{Zhang2020,Krizhevsky2012,Zhang2018} for
further details. 

\subsection{2-phases Learning}
To give a flavour of the algorithm imagine a dataset of points that depends only on one parameter, e.g. $\mu$, and the dataset is generated by sweeping the parameter $\mu$ inside of an interval $[A, B]$ (in case of multiple parameters, the same thing can be obtained by  keeping one fixed and varying the other one) as shown in Figure~\ref{fig:cflearning}(a).
Every element of the dataset is arranged into a matrix of shape depending on the observables making up the data point and is assigned to a label, $``0"$ or $``1"$.
The CNN is then trained in order to learn how to assign labels correctly. The degree of precision achieved by the network is evaluated by computing the \textit{accuracy}, which is defined as the ratio between the number of right guesses and the total number of guesses, over a test set.
It is important to stress that data belonging to the test set are not fed to the network during the learning process. 

In Learning by Confusion we select an interval like $[A,B]$ and a sweeping parameter $\mu_c$. Initially, we label the dataset uniformly, e.g. assign to each point the label $``0"$ corresponding to one of either phases (Fig.~\ref{fig:cflearning}(b)). The CNN is subsequently trained and tested (Fig.~\ref{fig:cflearning}(c)). As all points are assigned the same label the network learns easily to associate the label $``0"$ to every input, resulting in a perfect accuracy of 1.
Once that has been done, the dataset is relabeled: the first element of the dataset (the one obtained by setting e.g. $\mu_c = A$) is now assigned the label $``1"$, while the rest of the data points remain $``0"$. The CNN is then trained, validated and tested again, and the accuracy is expected to decrease because we are forcing the CNN to classify the dataset in a wrong way. We then proceed to relabel the data by assigning the first two data points to $``1"$ and the following ones to $``0"$. These steps represent the confusion part of the algorithm because we are deliberately mislabelling our dataset. The process is repeated until uniform $``1"$ labelling, and therefore again perfect accuracy is reached. We call this process the $2$--phases Learning by Confusion.

By plotting the accuracies obtained we encounter three possible scenarios. If the portion of the phase diagram belongs to the same phase we will see a characteristic V-shape with the lowest accuracy of 50\% being reached in the middle of the interval swept.

If the system undergoes a phase transition in the phase diagram region swept by the dataset, at a certain point during the confusion process the data will be correctly labelled according to the two phases. If that is the case, we expect a peak in the accuracy because the dataset is now labelled in a sensible way that the CNN can understand. This results in an \textit{accuracy function} characterized by the so-called W-shape of the accuracy plot~\cite{van_Nieuwenburg_2017}. 

Finally, if there is more than one phase transition it is not easy to predict the behavior of the accuracy function.

\subsection{3-phases learning}
Building on the intuition of the 2--phase learning, we considered an intuitive but yet unexplored extension of the model to detect two phase transitions. In this case, we have to label our dataset according to three labels instead of two, e.g. $``0"$, $``1"$ and $``2"$, and consider two possible transition points, $\mu^{(1)}_c, \ \mu^{(2)}_c$ that iteratively sweep through the interval of interest.

This operation results in an accuracy matrix in which each entry is the accuracy obtained by training the CNN setting the first phase transition point $\mu_c^{(1)}$ to its row number and the second one to its column number.
Each column represents a different value for the first transition $\mu_c^{(1)}$, and each row a different value for the second one $\mu_c^{(2)}$. In this way we obtain an accuracy that depends on two variables, that we represent via a contour plot. The plot is expected to display large accuracy values at the vertex points because it corresponds to the trivial to learn uniform $``0"$, $``1"$ and $``2"$ labellings. In addition to that, a maximum for the values of ($\mu_c^{(1)}$, $\mu_c^{(2)}$) corresponding to the two phase transitions is expected in the middle of the graph.
Results are clearly symmetrical w.r.t the diagonal since the transposition operation simply corresponds to an inversion between $``1"$ and $``2"$ labels.

It must be stressed that, contrarily to naive intuition, fixing one phase transition point, e.g. $\mu_c^{(1)}$, while sweeping the second one is not equivalent to performing 2-phases learning on a reduced dataset. In fact, one should keep in mind that the underlying CNN has been modified from a binary classifier to a ternary one; this naturally has non-trivial consequences on the output.

\section{Results}
\label{sec:Results}
We sum up the results of applying 2-phases and 3-phases learning to sections of the phase diagrams of four models: the Kitaev chain in its normal and interacting form and the one-dimensional Extended Hubbard model with two interaction ranges and fillings. For each model we briefly present its phase diagram and the results obtain with the confusion scheme, leaving the details of data and implementation to the appendix~\ref{sec:appendix}.

\subsection{Kitaev Chain}
\begin{figure*}[t!]
    \centering
    \includegraphics[width = \linewidth]{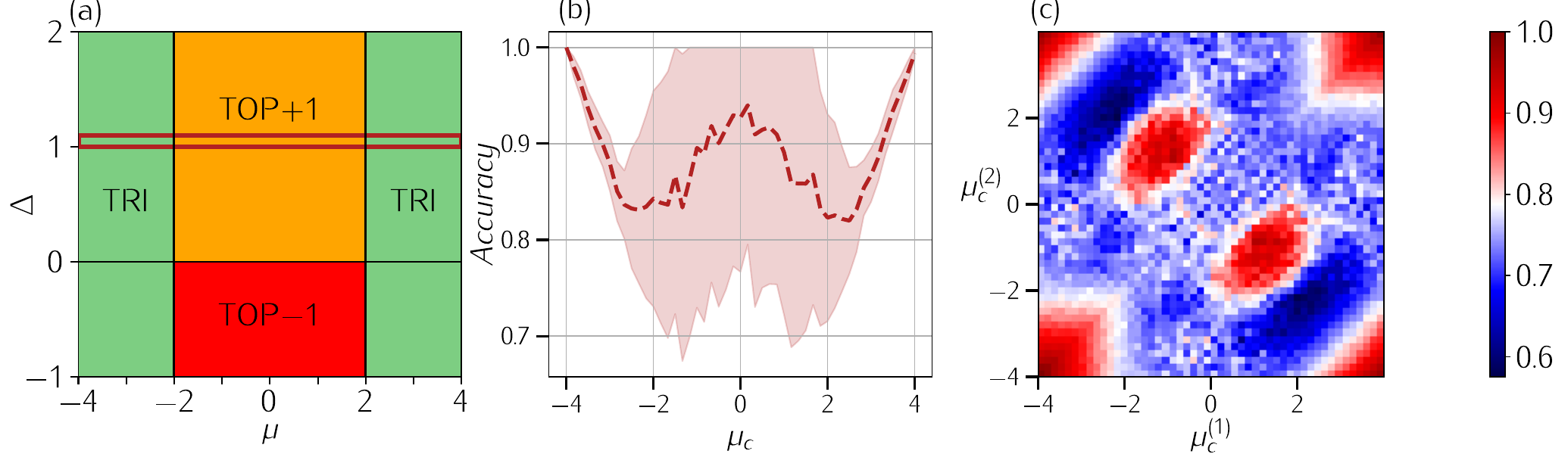}
    \\
    \includegraphics[width=\linewidth]{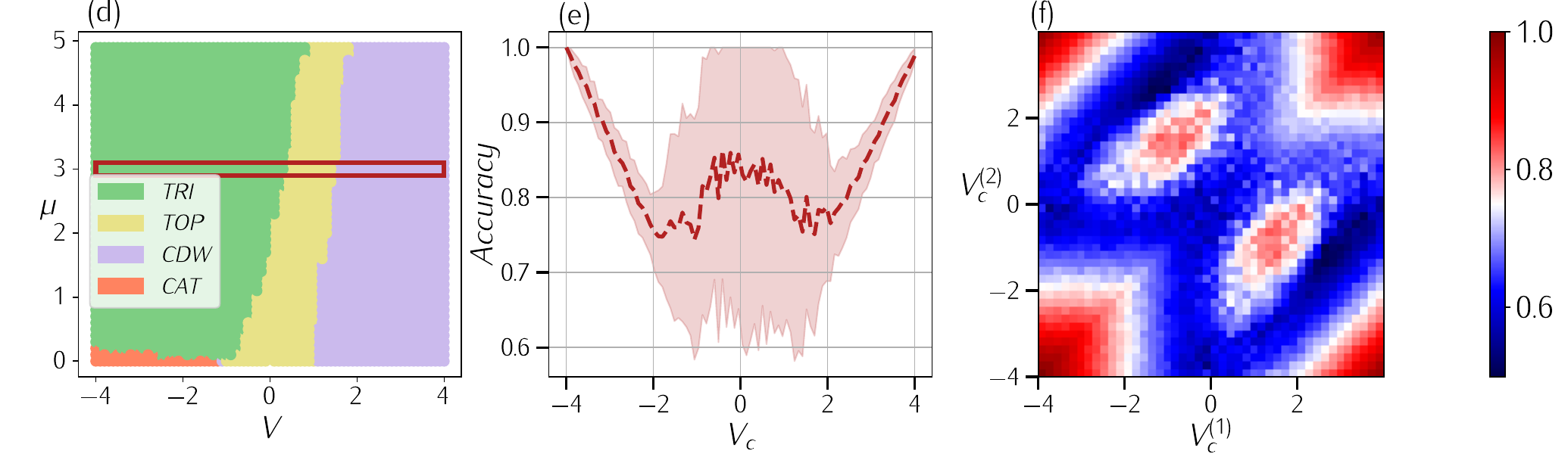}
    \caption{\textbf{2-phases and 3-phases learning on the Kitaev model}. The free Kitaev model: (a) Phase diagram  for $\mu \in [-4, 4], \Delta \in [-1, 2]$ presenting one trivial phase (TRI) and two topological phases (TOP$+1$/TOP$-1$) for $|\mu|  \geq 2\Delta$. In red, at $\Delta=1$ the line chosen to test the models.  (b) 2-phases learning applied to Kitaev. (c) 3-phases learning that predicts the two phase transitions of the line considered. Interacting Kitaev model: (d) Phase diagram, in red the section considered for confusion learning at $\mu = 3$. (e) 2-phase learning shows inconclusive results. (f) 3-phase learning shows a peak at two phase transition points $V_c^{(1)}, V_c^{(2)}$.}
    \label{fig:KitaevResults}
\end{figure*}
\subsubsection{Free Model}
The one--dimensional Kitaev chain~\cite{Kitaev2001} is a pedagogical model to show superconducting and topological effects. Given a chain of $L$ sites the Hamiltonian of the non--interacting (NI) version can be written as:
\begin{equation}
H^{\text{K}} =  \sum_{i=1}^L\big(J a^\dag_i a_{i+1} + \Delta \ a_{i} a_{i+1} + \text{h.c.} \big)  + \mu \sum_{i=1}^L  a^\dag_i a_i.
\label{eq:fermionic_hamiltonian}
\end{equation}
Here, $a_i \ (a_i^\dagger)$ creates (annihilates) spinless fermions on site $i$, $J$ is the nearest neighbor hopping coefficient, $\Delta$ is the superconducting pairing, $\mu$ is the chemical potential and we consider Periodic Boundary Conditions (PBC), i.e. $a_{L+1}\equiv a_1$. By going to momentum space and performing a Bogoliubov transformation we cast $H^{\text{K}}$ into a diagonal form $
H^{\text{K}}  = \sum_k E(k) \eta^\dag_k \eta_k, $ where $\eta_k$ are Bogoliubov operators and the single-particle energy $E(k)$ is given by
\begin{equation}
\label{eq:dispersionRelation}
E(k)  = \sqrt{h_z(k)^2 + h_y(k)^2},
\end{equation}
with 
\begin{gather}
\label{eq:hamiltonianVector}
h_z(k) =  J \cos k + \mu / 2,  \quad
h_y(k) = \Delta \sin k.
\end{gather}
This model describes a one-dimensional topological superconductor belonging to the BDI symmetry class~\cite{Schnyder2008,Chiu2016,Slager2013}, meaning that it presents time-reversal, particle-hole and chiral symmetry. In this symmetry class each topological phase is identified by a \textit{winding number}. This generates a phase diagram like the one shown in Fig~\ref{fig:KitaevResults} where we have topological phases with winding number $\pm 1$ (named TOP$+1$ and TOP$-1$) for $|\mu| < 2\Delta$ related to the presence of one Majorana mode at the edges of the chain. Otherwise, the phase is trivial (TRI).

We test 2-phases learning choosing $\Delta = 1$ and varying our trial transition point $\mu_c \in [-4, 4]$. The accuracy obtained is shown in Fig.~\ref{fig:KitaevResults}(b) and it does not respect the $\textit{W shape
}$ highlighting the possibility of 2 phase transitions. Therefore, we apply 3-phases learning with the two trial transition points $\mu_c^{(1)}, \mu_c^{(2)}$ varying in the same range and obtain the accuracy matrix shown in Fig.~\ref{fig:KitaevResults}(c). The central peak in accuracy is obtained for $(\mu_c^{(1)}, \mu_c^{(2)}) \sim (1,-1)$. 

\subsubsection{Interacting Model}
Adding an interaction term we obtain the Interacting Kitaev chain, already studied in~\cite{Stoudenmire2011, Hassler2012, Thomale2013, Katsura:2015tx, Miao:2017uf,Fromholz2020} 
\begin{equation}
H^{\text{K}}_{\text{int}} =  H^{\text{K}} + V \sum_i n_i n_{i+1}
\label{eq:fermionicHamiltonianInt}
\end{equation}
where $n_i=a^\dag_i a_i$ is the occupation number at site $i$. This model cannot be solved exactly due to the interacting potential. Therefore, we have reproduced the phase diagram by means of DMRG algorithm~\cite{Schollwock2005} with the ITensor package~\cite{itensor} after setting $J = \Delta = 1$ and we show it in Fig.~\ref{fig:KitaevResults}(d). The topological phase (TOP, yellow) was detected through the presence of a Majorana edge mode in the groundstate obtained with DMRG. The other phases (which do not present edge modes) are a Schrodinger's cat like phase (CAT, orange) and a Charge Density Wave phase (CDW, purple) that known in literature~\cite{Hassler2012,Thomale2013,Katsura:2015tx, Miao:2017uf} and a trivial phase (TRI, light green). 

We applied the confusion scheme to the line at $\mu=3$ with varying $V$. The plot in Fig~\ref{fig:KitaevResults}(b) shows the accuracy obtained by varying the transition point $V_c$ in the range $[-4,4]$. Since there is not only one phase transition we do not obtain the expected W-shape. Once again, the 3-phases learning plot has one peak in the middle, its coordinate being the two phase transition points. Although it is not exact, it predicts two phase transitions at $V_c^{(1)} \sim [1, 1.5]$ and $V_c^{(1)} \sim [0, -1]$.

\subsection{Extended Hubbard}
\begin{figure*}[t!]
    \centering
    \includegraphics[width=\linewidth]{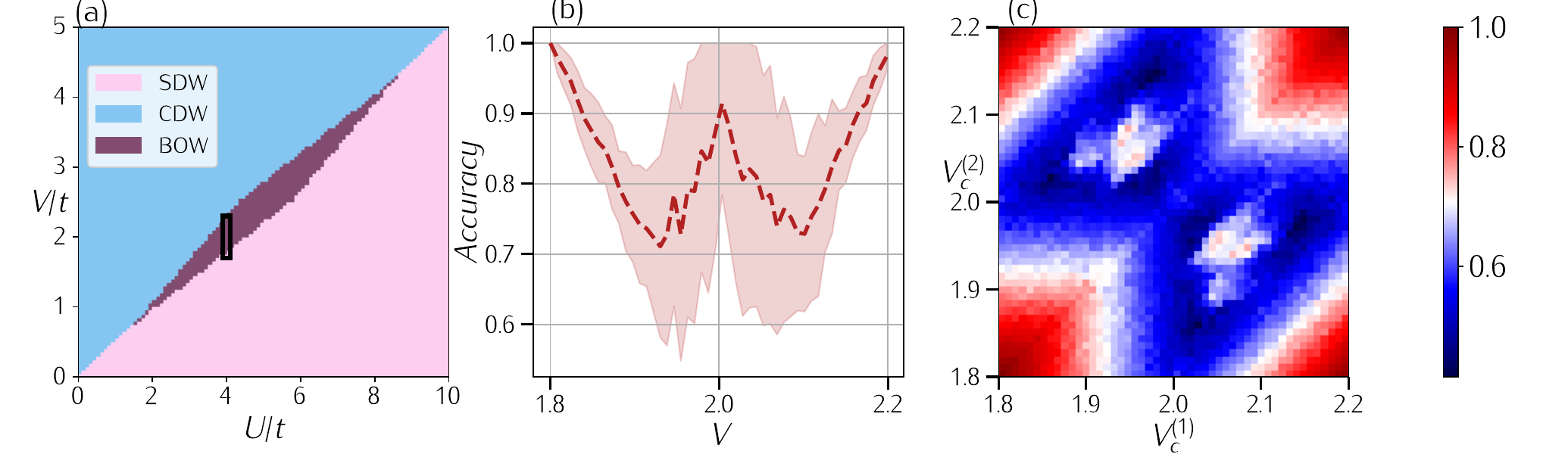}
    \\
    \includegraphics[width=\linewidth]{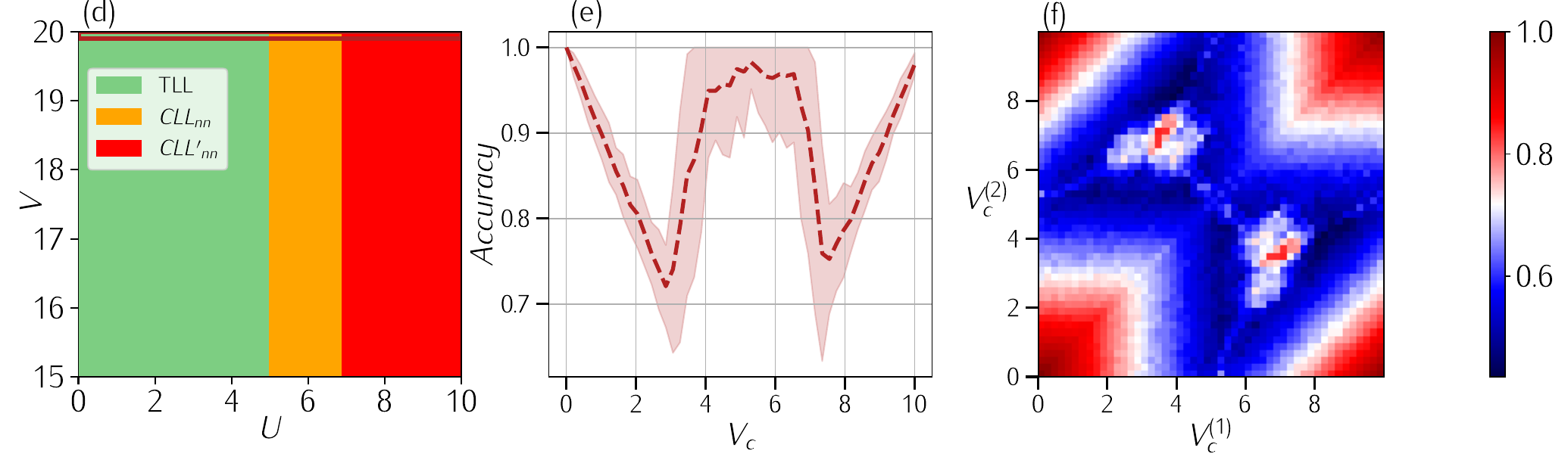}
    \caption{\textbf{2-phases and 3-phases learning applied to Extended Hubbard with } $\mathbf{r_c=1, 2}$. (a) Phase diagram of the $r_c=1 Model$ showing the CDW and SDW sectors separated by the thin BOW phase. The black rectangle indicates the points where confusion learning was applied. (b) For this model, 2-phase learning detects a single phase transitions at $V=2t$, while (c) 3-phase learning shows a peak at the two close phase transition points $V_c^{(1)} \simeq 1.9t$ and $V_c^{(2)} \simeq 2.1t$. (d) Phase diagram for high $U$ values of the $r_c=2$ Model with three phases, the region investigated with confusion learning is highlighted by the red rectangle. In this case: (e) 2-phases learning returns a plateau of high accuracy for all the values inside the $CLL_{nn}$ phase, while (f) 3-phases learning shows a clear peak in accuracy at coordinates $V^{(1)}_c\simeq3.8t$, $V^{(2)}_c\simeq6.8t$, in accordance with previous results.}
    \label{fig:HubbardResults}
\end{figure*}
The fermionic
Hubbard model~\cite{Arovas_2022,tasaki1997hubbard} has been deeply studied in the past due to its solvability in one dimension, which can give physical insights regarding strongly correlated electronic systems, and due to the growing possibilities of simulating it through quantum technologies
\cite{PhysRevLett.126.210601,Hensgens:2017,tarruell2019quantum}.
In recent years extensions of this model have been interest of study in order to investigate high temperature superconductivity
\cite{Robaszkiewicz_1999,Dong_2022}. 

In the following we will consider the one-dimensional Hamiltonian 
\cite{PhysRevB.61.16377}
\begin{align}
\begin{split}
    H = &- t \sum_{i,\sigma=\uparrow,\downarrow}
    \bigg( a^\dag_{i+1,\sigma}  \ 
    a_{i,\sigma} + \text{h.c.} \bigg) \\
    &+  U  \sum_i n_{i\uparrow} n_{i\downarrow} +
    V \sum_i \sum_{l=1}^{r_c} n_i n_{i+l}
\label{eq:HubbardHamiltonian}
\end{split}
\end{align}
where $a^\dag_{i,\sigma}$, $a_{i,\sigma}$ are the usual fermionic creation and annihilation operators for particles of spin
$\sigma=\uparrow,\downarrow$ , 
$n_{i\sigma}=a^\dag_{i,\sigma}a_{i,\sigma}$ and 
$n_i = n_{i\uparrow} + n_{i\downarrow}$. The Hamiltonian~\eqref{eq:HubbardHamiltonian} represents the famous Hubbard model, which includes nearest-neighbors hopping and on-site interaction, with the addition of a soft-shoulder interaction term with range of $r_c$ sites. In particular, we refer to $U,\ V$ for the on-site and off-site interaction strengths, respectively. As the Hamiltonian~\eqref{eq:HubbardHamiltonian} commutes with the total number of particles $N=N_\uparrow + N_\downarrow$, $N_{\sigma}=\sum_i n_{i,\sigma}$, the nature of the ground state is also dictated by the filling; this is defined as $\rho= \rho_\uparrow + \rho_\downarrow$, with $\rho_{\sigma}=N_\sigma/2L$.

\subsubsection{\texorpdfstring{$r_c = 1$ Model}{}}

Taking $r_c =1$~\cite{Satoshi2007} at half-filling $\rho_\uparrow=\rho_\downarrow=1/4$, i.e. one particle per site, one gets a relatively simple phase diagram, as shown in Fig.~\ref{fig:HubbardResults}(a). Here, between the usual Charge and Spin Density Wave phases (CDW, light blue/SDW, pink), which are classically separated by the phase transition line for $U/t = 2 V/t$, there is a small region exhibiting  a Bond Order Wave (BOW, purple) phase. In particular, this should be present near the classical phase transition line and for $V/t\in[1,4]$~\cite{Satoshi2007}.
To test the effectiveness of our construction onto this model we generated a set of data with fixed value of $U/t = 4$ and sweeping $V/t$ in an interval $[1.8, 2.2]$ restricted close to the phase transition line. Although 2-phases learning detects a phase transition precisely at $V/t = 2$ (panel (b) of Fig.~\ref{fig:HubbardResults}), 3-phases learning (panel (c)) detects the possibility of two phase transitions happening at values $V_c^{(1)} \simeq 1.9t$ and $V_c^{(2)} \simeq 2.1t$. This is in agreement with results present in the literature, as these transitions are predicted to happen, in the thermodynamic limit and for $U=4t$, at $V_c^{(1)}\simeq1.877t$ and $V_c^{(2)}=2.164t$. The ability to catch this transition should not be taken for granted, as the transition from CDW to BOW has been object of many analytical \cite{PhysRevLett.53.2327,PRB_1990,PhysRevB.44.3559,PhysRevB.49.7904,JVoit_1995,PhysRevLett.88.056402,PhysRevLett.96.036408} and numerical \cite{PhysRevLett.89.236401,PhysRevB.65.155113,PhysRevLett.92.236401,PhysRevLett.92.246404,Glocke_2007} investigations in the past because of the difficulty in its detection. This could be the reason why both 2-phase and 3-phase learning (Fig.~\ref{fig:HubbardResults} (b) and (c)) identify this phase transitions. The aforementioned difficulty is probably the reason why both the 2-phases and 3-phases schemes show satisfactory results. Nonetheless, the ability of the 3-phases extension of Learning by Confusion to detect this non-trivial phase transition signals once again the adaptability and power of our construction.

\subsubsection{\texorpdfstring{$r_c = 2$ Model}{}}
We now turn our attention to another extension of the model with the same Hamiltonian as Eq.~\ref{eq:HubbardHamiltonian} but with $r_c = 2$ and filling $\rho=2/5$. In this case we consider a chain of $L=30$ sites with PBC at filling $\rho=2/5$, with $\rho_\uparrow=\rho_\downarrow=1/5$. The choice of $L=30$ is connected to the frustration of the model for these particular values of $r_c$ and $\rho$. A full description of the phase diagram of this model is way beyond the purposes of this introductory section and we address the interested reader to~\cite{https://doi.org/10.48550/arxiv.1909.12168,Botzung_thesis,ourpaper}. As we consider the strong on-site interaction limit $U\gg t$ (in particular we take $U/t=20$), the ground state of the model should be similar to the one studied in \cite{Mattioli_2013,Dalmonte_2015} for a spinless chain. In particular, for the same filling $\rho=2/5$, it was predicted in~\cite{Dalmonte_2015} that a phase should emerge for $V\in[4t,6t]$, separating a Tomonaga Luttinger Liquid phase to a Cluster Luttinger Liquid one. By sweeping  $V/t\in[0,10]$ we find that, while $2$-phases learning scheme does not display the desired W-shaped accuracy, the $3$-phases construction presents a peak for $V^{(1)}_c\simeq3.8t$, $V^{(2)}_c\simeq6.8t$, in good agreement with literature results.

\section{Conclusions}
\label{sec:Conclusions}

In this work, we proposed a generalization of the Learning by Confusion scheme, a technique which has proved useful in inspecting quantum many-body equilibrium phase diagrams in an unbiased way. While the original scheme was mostly designed to address systems exhibiting only one transition point, we tackled the problem of phase diagrams regions with three phases. Our construction simply addresses the problem by modifying the underlying Convolutional Neural Network from a binary classificator to a ternary one, paving the way to systems with an arbitrary number of phases. Moreover, we tested this construction onto a variety of different models, ranging from the integrable case of the free Kitaev chain and its non-integrable interacting extension to the one-dimensional Extended Hubbard model with soft-shoulder potential. Despite the different natures of transition and phases displayed by the aforementioned models, the $3$-phases confusion was always able to identify phase transition points consistent with existing results in the literature.
Besides the immediate proof of principle regarding the technique itself, our work paves the way to extending Learning by Confusion to a variety of new systems, showing once more the usefulness of Neural Networks within the context of quantum many-body theory.

\section*{ACKNOWLDGEMENTS}
F.C. acknowledges support  by PEPR-Q (QubitAF project). Numerical simulations have been performed on the PSMN cluster at the ENS of Lyon and on INFN Bologna clusters. The research of S.T. and E.E is partially funded by the International Foundation Big Data and Artificial Intelligence for Human Development (IFAB, project “Quantum Computing for Applications”) and by INFN (project “QUANTUM”). E.E. also acknowledges financial support from the National Centre for HPC, Big Data and Quantum Computing (Spoke 10, CN00000013)
\bibliography{bibliography.bib}

\appendix

\section{Data and Details}
\label{sec:appendix}
For each model we produced a set of observable at different points of the phase diagram to create a dataset. Here we list the specific for each model.

\subsection{Kitaev chain} We define the Fourier transform of the single-particle standard ($c(k)$) and anomalous ($f(k)$) correlation functions:
\begin{gather}
c(k) =\sum_{i, j} e^{\mathrm{i} k (i - j)} \langle a^\dag_{i} a_{j} \rangle\label{eq:ck_corr} \\
f(k) = \sum_{i, j} e^{\mathrm{i} k (i - j)} \langle a_{i} a_{j} \rangle\label{eq:fk_corr}
\end{gather}
where the expectation values are taken over the ground state. Notice that $c(k)$ is real whereas $f(k)$ is purely imaginary. This is due to the antisymmetry of the expectation value $ \langle a_{i} a_{j} \rangle$ for the exchange $i \leftrightarrow j$. For this reason we only take the imaginary part of $f(k)$. In our non--interacting model (Eq.~\eqref{eq:fermionic_hamiltonian}), the correlators $c(k)$ and $f(k)$ can be computed analytically and take the form:
\begin{gather} 
	c(k) = \frac{1}{2} + 
	\frac{\mu/2 + J\cos k}{2 {E(k)}},
 \label{corrfun1}\\
	f(k) = \frac{\Delta \sin k}{2 {E(k)}},
	\label{corrfun2}
\end{gather}
with $E(k)$ being the energy dispersion relation~\ref{eq:dispersionRelation} in the text. We created a dataset of 1500 points with $\mu=1$ fixed and varying $\Delta$ in the range $[-4, 4]$ making sure to cover each phase transition. Each point is a $2\times L$ matrix with the two correlators stacked to be fed into the convolutional neural network. Then, the dataset was divided into 50 ordered subsets according to their $\Delta$ to perform the confusion scheme. 

In the interacting case, it is not possible to evaluate exactly the correlation functions $c(k)$ (Eq.~\eqref{eq:ck_corr}) and $f(k)$ (Eq.~\eqref{eq:fk_corr}) on the ground state of the Hamiltonian of Eq.~\eqref{eq:fermionicHamiltonianInt}. Therefore we calculate them by means of the DMRG algorithm for a lattice of size $L=100$. We generate 4000 points for parameters $V \in [-4, 4]$ and $\mu \in [0, 5]$ being the phase diagram symmetrical w.r.t. the transformation $\mu \rightarrow -\mu$. Each point is a $2\times 100$ matrix, the first row being the correlator $c(k)$ (Eq.~\eqref{eq:ck_corr}) and the second being $f(k)$ (Eq.~\eqref{eq:fk_corr}). In the same way as the Kitaev chain, we created a dataset of all the points with $\mu=3$, ordered them according to their $V$ and divided them in 50 subgroups to perform the confusion scheme.

\subsection{Hubbard Model}
In the case of the one-dimensional extended Hubbard model with soft-shoulder potential with $r_c=1$ at half-filling, previous works have shown the presence of three phases for $U=4t$ and $V\in[1.8t,2.2t]$. These are for increasing values of $V$, Charge Density Wave, Bond-Order Wave and Spin Density Wave. For this reason, we identified three observables which are capable of detecting these phases. For CDW and SDW we take the charge and spin structure factors, defined as
\begin{align}
    & S_c(k) = \frac{1}{N} \sum_{l,j} e^{-ik(j-l)} \bigg( \braket{n_l n_j} - \braket{n_l}\braket{n_j} \bigg) \\
    & S_s(k) = \frac{1}{N} \sum_{l,j} e^{-ik(j-l)} \bigg( \braket{S^z_l S^z_j} - \braket{S^z_l}\braket{S^z_j} \bigg)
\end{align}
and the BOW order parameter, defined as
\begin{equation}
    B_i = \braket{c^\dag_i c_{i+1} + c^\dag_{i+1}c_i}.
\end{equation}

For the one-dimensional extended Hubbard model with soft shoulder potential with range $r_c=2$ at filling $\rho=2/5$ we consider again the charge and spin structure factors and the local density, defined as
\begin{equation}
   n(x_i) = \braket{n_i}. 
\end{equation}
\paragraph{}
\end{document}